\documentclass[amssymb,useAMS,prd,aps,amsmath,superscriptaddress,nofootinbib,twocolumn,preprintnumbers]{revtex4}
\pdfoutput=1

\usepackage{color}
\usepackage{pslatex}
\usepackage{pdfpages}
\usepackage{graphicx}
\usepackage{psfrag}
\usepackage{pdftricks}
\usepackage{multirow}
\usepackage{graphicx}
\usepackage[none]{hyphenat}
\usepackage{url}
\usepackage[breaklinks, colorlinks, citecolor=blue]{hyperref}
\usepackage{ifthen}
\usepackage{ulem}
\def\eprinttmp@#1arXiv:#2 [#3]#4@{
\ifthenelse{\equal{#3}{x}}{\href{http://arxiv.org/abs/#1}{#1}}{\href{http://arxiv.org/abs/#2}{arXiv:#2} [#3]}}
\providecommand{\eprint}[1]{\eprinttmp@#1arXiv: [x]@}
\newcommand{\adsurl}[1]{\href{#1}{ADS}}

\begin{document}

\title{Using the CMB angular power spectrum to study Dark Matter--photon interactions}

\author{Ryan J. Wilkinson}
\affiliation{Institute for Particle Physics Phenomenology, Durham University, Durham, DH1 3LE, United Kingdom}

\author{Julien Lesgourgues}
\affiliation{Institut de Th\'eorie des Ph\'enom\`enes Physiques, \'Ecole Polytechnique F\'ed\'erale de Lausanne, CH-1015, Lausanne, Switzerland}
\affiliation{CERN, Theory Division, CH-1211 Geneva 23, Switzerland}
\affiliation{LAPTH, U. de Savoie, CNRS,  BP 110, 74941 Annecy-Le-Vieux, France}

\author{C\'eline B\oe hm}
\affiliation{Institute for Particle Physics Phenomenology, Durham University, Durham, DH1 3LE, United Kingdom}
\affiliation{LAPTH, U. de Savoie, CNRS,  BP 110, 74941 Annecy-Le-Vieux, France}

%%%%%%%%%%%%%%%%%%%%%%%%%%%%%%%%%%%%%%%%%%%%%%%%%%%%%
%%%%%%%%%%%%%%%%%%%%%%%%%%%%%%%%%%%%%%%%%%%%%%%%%%%%%

\begin{abstract}
In this paper, we explore the impact of Dark Matter--photon interactions on the CMB angular power spectrum. Using the one-year data release of the Planck satellite, we derive an upper bound on the Dark Matter--photon elastic scattering cross section of $\sigma_{\rm{DM}-\gamma} \leq 8 \times 10^{-31} \left(m_{\rm{DM}}/\rm{GeV}\right) \ \rm{cm^2}$ (68\% CL) if the cross section is constant and a present-day value of $\sigma_{\rm{DM}-\gamma} \leq 6 \times 10^{-40} \left(m_{\rm{DM}}/\rm{GeV}\right) \ \rm{cm^2}$ (68\% CL) if it scales as the temperature squared. For such a limiting cross section, both the $B$-modes and the $TT$ angular power spectrum are suppressed with respect to $\Lambda$CDM predictions for $\ell \gtrsim 500$ and $\ell \gtrsim 3000$ respectively, indicating that forthcoming data from CMB polarisation experiments and Planck could help to constrain and characterise the physics of the dark sector. This essentially initiates a new type of dark matter search that is independent of whether dark matter is annihilating, decaying or asymmetric. Thus, any CMB experiment with the ability to measure the temperature and/or polarisation power spectra at high $\ell$ should be able to investigate the potential interactions of dark matter and contribute to our fundamental understanding of its nature.
\end{abstract}\preprint{IPPP/13/59, DCPT/13/118, CERN-PH-TH-2013-208, LAPTH-046/13}

\date{\today}

\maketitle

%%%%%%%%%%%%%%%%%%%%%%%%%%%%%%%%%%%%%%%%%%%%%%%%%%%%%
%%%%%%%%%%%%%%%%%%%%%%%%%%%%%%%%%%%%%%%%%%%%%%%%%%%%%

%%%%%%%%%%%%%%%%%%%%%%%%%%%%%%%%%%%%%%%%%%%%%%%%%%%%%
\section{Introduction}
\label{sec:intro}
%%%%%%%%%%%%%%%%%%%%%%%%%%%%%%%%%%%%%%%%%%%%%%%%%%%%%

The last decade has witnessed tremendous progress in observational cosmology. From the accumulated data of e.g. supernovae surveys~\cite{Kowalski:2008ez}, BAO measurements~\cite{Eisenstein:2005su} and Cosmic Microwave Background (CMB) experiments such as WMAP~\cite{Hinshaw:2012aka},  SPT~\cite{Hou:2012xq}, ACT~\cite{Sievers:2013ica} and more recently Planck~\cite{Ade:2013ktc}, one could establish with great precision the quantity of dark matter (DM) in the Universe. Yet, despite the large number of dedicated experiments, an understanding of the particle nature of DM and direct evidence for its existence have remained elusive, questioning our interpretation of this mysterious substance.

One of the possible explanations for this long-standing puzzle is that DM consists of weakly-interacting massive particles (WIMPs) that are naturally difficult to detect using methods based on their interactions. Quantitative estimates of the DM interaction rate have been made using direct~\cite{Ahmed:2010wy,Aprile:2012nq} and indirect~\cite{Adriani:2008zr,Abbasi:2009uz,Ackermann:2012qk} detection techniques, in addition to relic density calculations~\cite{Gondolo:1990dk}. Such methods generally assume that DM annihilates (in our galaxy and including at late times) and/or interacts with quarks.

However, such assumptions are not always appropriate; for instance, there are no late annihilations in asymmetric DM scenarios that could lead to a visible signal in galactic or CMB data~\cite{Petraki:2013wwa}. Additionally, the DM mass may be too small or too large to produce a visible signal in direct detection experiments due to their limited sensitivity\footnote{However, new techniques are now being proposed to probe the lighter mass range, see e.g. Ref.~\cite{Barreto:2011zu}.}. For example, if DM consists of sterile neutrinos (e.g.~\cite{Dodelson:1993je,Boyarsky:2009ix,Viel:2005qj,Dolgov:2000ew}) with a significant decay rate, X-ray observations~\cite{Fuller:2003gy,Boyarsky:2005us,Watson:2006qb,Abazajian:2006jc} would be a more promising detection method than direct detection. Also, if DM is lighter than $\sim$ 10 GeV~\cite{Boehm:2002yz} with a small annihilation cross section into electron--positron pairs~\cite{Boehm:2002yz,Boehm:2003bt}, it is more appropriate to look for evidence in low energy gamma-ray data~\cite{Boehm:2003hm}, measurements of the electron/muon g-2~\cite{Boehm:2004gt,Boehm:2007na,Hanneke:2010au,Bouchendira:2010es} or the neutrino mass generation mechanism~\cite{Boehm:2006mi}. However, such searches require one to assume a Particle Physics model and are therefore not universal. Finally, DM could be much heavier than a few TeV (e.g.~\cite{Bergstrom:1994mg}), posing problems for the usual detection techniques.

Here we propose an alternative method to determine how weak DM interactions with Standard Model (SM) particles need to be, independently of the vanilla DM assumptions. Our argument holds whether DM decays, annihilates or is in the right mass range to interact significantly with nuclei. It is only based on the historical motivation for WIMPs, namely the mandatory absence of Silk damping (photon diffusion) at very large scales~\cite{Boehm:2000gq,Boehm:2001hm,Boehm:2004th}.

To begin with, it is important to note that charge neutrality does not necessarily rule out Dark Matter--photon (DM--$\gamma$) interactions since they can occur through more complicated processes involving SM particles\footnote{Recent arguments for limited electromagnetic interactions can be found in Ref.~\cite{McDermott:2010pa}.} or magnetic and dipole moments~\cite{Sigurdson:2004zp,Gardner:2008yn,Chang:2010en}. Therefore, in principle, DM could have couplings with a strength intermediate between those of the electromagnetic and (SM) weak interactions. From a phenomenological point of view, the prejudice is that these interactions should be relatively small\footnote{For example, in Supersymmetry, the neutralino pair annihilation cross section into two photons is expected to be smaller than $10^{-38}~\rm{cm^2}$~\cite{Bergstrom:1997fh}. However, in some conditions (in particular when the neutralino is a wino and is thus mass degenerated with a chargino), the annihilation cross section can be much larger than $10^{-34}~\rm{cm^2}$~\cite{Hisano:2002fk}. Whether such large values of the annihilation cross section translate into large values of the elastic scattering cross section is beyond the scope of this paper. Here we will simply assume that there are realistic DM scenarios where the DM--$\gamma$ elastic scattering cross section is significant.} but since we lack evidence in favour of any particular DM model, deriving constraints from the accumulated cosmological data offers a more robust method to characterise its nature.

The consequence of DM interactions with SM particles is to dampen the primordial matter fluctuations and essentially erase all structures below a given scale (referred to as the collisional damping scale)~\cite{Boehm:2000gq,Boehm:2001hm,Boehm:2004th}. The effect is exacerbated when DM couples to photons and therefore, one can set a strong upper limit on the DM--$\gamma$ interaction cross section by examining the resulting CMB spectra.

In fact, a non-zero $\rm{DM}-\gamma$ coupling has two specific signatures. Firstly, as was shown in Ref.~\cite{Boehm:2001hm}, large interactions lead to the presence of significant damping in the angular power spectrum, which can be constrained using the position and relative amplitude of the acoustic peaks. Secondly, after DM ceases to interact with photons, the collisional damping is supplemented by DM free-streaming\footnote{Assuming the DM--$\gamma$ decoupling happens before the gravitational collapse of such fluctuations and the DM velocity is not completely negligible at this time; this offers a way to determine the decoupling epoch.}; this appears as a `linear' translation of the matter power spectrum and can also be constrained (if the effect is substantial enough). Therefore, with the first data from the Planck satellite~\cite{Ade:2013zuv}, one can set a limit on DM--$\gamma$ interactions with unprecedented precision.

In this study, we extend the preliminary analysis of Ref.~\cite{Boehm:2001hm} much further and show that a non-negligible DM--$\gamma$ coupling also generates distinctive features in the temperature and polarisation power spectra at high $\ell$. One can use these effects to search for evidence of DM interactions in CMB data and determine (at least observationally) the strength of DM--$\gamma$ interactions that we are allowed. This work will be extended to other DM interactions in a future publication.

The paper is organised as follows. In Sec.~\ref{sec:imp}, we discuss the implementation of DM--$\gamma$ interactions and the qualitative effects on the $TT$ and $EE$ components of the angular power spectrum. In Sec.~\ref{subsec:results}, we constrain these interactions by comparing the spectra to the latest Planck data, and find the best-fit cosmological parameters. In Sec.~\ref{subsec:predictions}, we present our predictions for the temperature and polarisation spectra for the maximally allowed value of the elastic scattering cross section that we obtain. We conclude in Sec.~\ref{sec:conc}.

%%%%%%%%%%%%%%%%%%%%%%%%%%%%%%%%%%%%%%%%%%%%%%%%%%%%%
\section{Implementation of the DM--$\gamma$ interactions}
\label{sec:imp}
%%%%%%%%%%%%%%%%%%%%%%%%%%%%%%%%%%%%%%%%%%%%%%%%%%%%%

In this section, we recall the modified Boltzmann equations used to incorporate interactions of DM with photons~\cite{Boehm:2001hm} and discuss their implementation in the Cosmic Linear Anisotropy Solving System ({\sc class}) code\footnote{\tt class-code.net} (version 1.7)~\cite{Lesgourgues:2011re,Blas:2011rf}. 

The current version of {\sc class} offers a choice between two gauges for the definition of cosmological perturbations: the Newtonian gauge, and the synchronous gauge comoving with DM (see e.g. Ref.~\cite{Ma:1995ey}). In the presence of coupled DM, the synchronous gauge equations should be slightly reformulated since the gauge can be fixed by imposing $\theta_{\rm{DM}}=0$ at the initial time but not at all times. For simplicity, we implemented the DM--$\gamma$ interactions in the Newtonian gauge only. All equations in this section refer to that gauge, assuming a flat universe and taking derivatives with respect to conformal time, $\tau$. Our notation is consistent with Ref.~\cite{Ma:1995ey}.

%%%%%%%%%%%%%%%%%%%%%%%%%%%%%%%%%%%%%%%%%%%%%%%%%%%%%
\subsection{Modified Boltzmann equations}
\label{subsec:bolt}
%%%%%%%%%%%%%%%%%%%%%%%%%%%%%%%%%%%%%%%%%%%%%%%%%%%%%

In the absence of DM interactions, the Boltzmann equations simplify to the following Euler equations:
\begin{eqnarray}
\label{vbar}
\dot \theta_{\rm b} 
 & = &   k^2 \psi - {\cal H} \theta_{\rm b} + c_s^2 k^2 \delta_{\rm b}
       - R^{- 1} \dot \kappa (\theta_{\rm b} - \theta_\gamma)~, \\
\label{vphot}
\dot \theta_\gamma
 & = &   k^2 \psi + k^2 \left(\frac{1}{4} \delta_\gamma - \sigma_\gamma\right) - \dot \kappa (\theta_\gamma - \theta_{\rm b})~, \\
\label{euler_DM}
\label{vcdm}
\dot \theta_{\rm{DM}}
 & = &   k ^2\psi - {\cal H} \theta_{\rm{DM}}~,
\end{eqnarray}
where $\theta_{\rm b}$, $\theta_\gamma$ and $\theta_{\rm{DM}}$ are the baryon, photon and DM velocity divergences respectively. $\delta_\gamma$ and $\sigma_\gamma$ are the density fluctuation and anisotropic stress potential associated with the photon fluid, $\psi$ is the gravitational potential, $k$ is the comoving wavenumber, ${\cal H}=(\dot a / a)$ is the conformal Hubble rate, $R \equiv (3/4) (\rho_{\rm{b}}/\rho_\gamma)$ is the ratio of the baryon to photon density, $c_s$ is the baryon sound speed and $\dot{\kappa} \equiv a \hspace{0.3ex}\sigma_{\rm Th}\hspace{0.3ex}c\hspace{0.3ex}n_e$ is the Thomson scattering rate (the scale factor, $a$, appears since the derivative is taken with respect to conformal time).

DM--$\gamma$ interactions are accounted for by a term analogous to $-\dot{\kappa} (\theta_\gamma - \theta_{\rm b})$ in the DM and photon velocity equations. The new interaction rate reads 
$\dot{\mu} \equiv a\hspace{0.3ex}\sigma_{\rm{DM}-\gamma}\hspace{0.3ex}c\hspace{0.3ex}n_{\rm{DM}}$, where $\sigma_{\rm{DM}-\gamma}$ is the DM--$\gamma$ elastic scattering cross section, $n_{\rm{DM}} = \rho_{\rm{DM}} / m_{\rm{DM}}$ is the DM number density, $\rho_{\rm{DM}}$ is the DM energy density and $m_{\rm{DM}}$ is the DM mass (assuming that DM is non-relativistic)\footnote{Intuitively, one can understand why $\dot{\mu}$ must be proportional to the cross section and the DM number density; if either the number of DM particles or the cross section is completely negligible, the photon fluid will not be significantly modified by a DM--$\gamma$ coupling.}. Thus, the Euler equation for photons receives the additional source term $- \dot \mu (\theta_\gamma - \theta_{\rm{DM}})$.

In order to conserve energy and account for the momentum transfer in an elastic scattering process, the source term in the Euler equation for DM has the opposite sign and is rescaled by a factor $S \equiv (3/4)(\rho_{\rm{DM}}/\rho_\gamma)$, which grows in proportion to $a$. Thus, the Euler equations become
\begin{eqnarray}
\label{vbar2}
\dot \theta_{\rm b} 
 & = &   k^2 \psi - {\cal H} \theta_{\rm b} + c_s^2 k^2 \delta_{\rm b}
       - R^{- 1} \dot \kappa (\theta_{\rm b} - \theta_\gamma)~, \\
\nonumber
\dot \theta_\gamma
 & = &   k^2 \psi + k^2 \left(\frac{1}{4} \delta_\gamma - \sigma_\gamma\right) \\
\label{vphot2}
 &&- \dot \kappa (\theta_\gamma - \theta_{\rm b}) - \dot \mu (\theta_\gamma - \theta_{\rm DM})~, \\
\label{euler_DM2}
\label{vcdm2}
\dot \theta_{\rm{DM}}
 & = &   k ^2\psi - {\cal H} \theta_{\rm{DM}} - S^{- 1} \dot \mu (\theta_{\rm{DM}} - \theta_\gamma)~.
\end{eqnarray}

The DM--$\gamma$ elastic scattering cross section, $\sigma_{\rm{DM}-\gamma}$, can be either constant (like the Thomson scattering between photons and charged particles) or proportional to temperature, depending on the DM model that is being considered.

For a constant cross section, since DM and baryons are non-relativistic when we begin the integration, both $\dot \mu$ and $\dot \kappa$ behave as $a^{- 2}$ at high redshifts. Therefore, the ratio of $\dot \mu$ and $\dot \kappa$ is proportional to the dimensionless quantity  
\begin{equation}
\label{u_gcdm}
u \equiv \left[\frac{\sigma_{\rm{DM}-\gamma}}{\sigma_{\rm Th}} \right] \left[\frac{m_{\rm{DM}}}{100~\rm{GeV}} \right]^{- 1}~,
\end{equation}
which depends on two essential parameters: the scattering cross section, $\sigma_{\rm{DM}-\gamma}$, and the DM mass, $m_{\rm{DM}}$\footnote{Note that after recombination, $\dot \kappa$ is strongly suppressed (by a factor $\sim 10^{-4}$~\cite{Peebles:1968ja}) due to the drastic subsequent drop in the free electron density, while $\dot \mu$ continues scaling like $a^{- 2}$.}. We will use this parameter to quantify the effect of DM--$\gamma$ interactions on the evolution of primordial fluctuations. If instead the cross section is proportional to the temperature squared (e.g. dipole DM~\cite{Sigurdson:2004zp,Gardner:2008yn,Chang:2010en} or by analogy to neutrino--electron scattering), we can write $u = u_0~a^{-2}$, where $u_0$ is the present-day value.

As the magnitude of the $u$ parameter determines the collisional damping scale~\cite{Boehm:2001hm}, one can readily see that the efficiency of the damping is essentially governed by the ratio of the interaction cross section to the DM mass.

%%%%%%%%%%%%%%%%%%%%%%%%%%%%%%%%%%%%%%%%%%%%%%%%%%%%%
\subsection{Implementation in {\sc class}}
\label{subsec:class}
%%%%%%%%%%%%%%%%%%%%%%%%%%%%%%%%%%%%%%%%%%%%%%%%%%%%%

The execution of {\sc class} begins by using three distinct modules for the background, thermodynamical and perturbation evolutions. In the present study, all necessary modifications are confined to the thermodynamics and perturbation modules. 

The standard thermodynamics module solves the recombination equations and stores an interpolation table for 
[$\dot \kappa$, $\ddot \kappa$, $\dddot \kappa$,  $\exp(-\kappa)$] as a function of redshift, $z$. At the same time, we request that the module stores the corresponding values of $\dot \mu$ (inferred analytically from $u$, $a$, $\sigma_{\rm Th}$ and $\rho_{\rm DM}$), its higher derivatives and $\exp(-\mu)$. It also stores values of the visibility function
\begin{equation}
\label{vis}
g(\tau)=(\dot \kappa + \dot \mu) e^{-\kappa-\mu}~,
\end{equation}
along with its first and second time derivatives.

In the perturbation module, we began by adding the new interaction terms to the photon and DM Euler equations [see Eqns.~\eqref{vphot2} and~\eqref{vcdm2}] and in the full hierarchy of Boltzmann equations for photon temperature and polarisation. Apart from the source term in the photon velocity equation, this amounts to simply replacing all occurrences of $\dot \kappa$ with $(\dot \kappa + \dot \mu)$. For instance, the evolution equation for photon temperature multipoles with $\ell \geq 3$ reads
\begin{equation}
\label{large_l_evolution}
\dot F_{\gamma \ell} = \frac{k}{2\ell+1} \left[ \ell F_{\gamma(\ell-1)} - (\ell+1) F_{\gamma (\ell+1)}\right]
- (\dot \kappa + \dot \mu) F_{\gamma \ell}~,
\end{equation}
where $F_{\gamma \ell}$ is defined as in Ref.~\cite{Ma:1995ey}.

At early times, the characteristic scale $\tau_c=(\dot \kappa)^{-1}$ is very small, leading to a stiff system of equations. Integrating over time remains efficient in the baryon--photon tight-coupling regime (in which small quantities like $(\dot \theta_\gamma- \dot \theta_{\rm b})$ and $\sigma_\gamma$ are obtained analytically at order one or two in the expansion parameter), while the remaining evolution equations become independent of $\tau_c$.

To obtain a CMB spectrum compatible with large-scale observations, we can limit our analysis to the case in which the new interaction rate is weaker than the Thomson scattering rate, i.e. $\dot \mu < \dot \kappa$. Therefore, there is no need to devise a specific DM--$\gamma$ tight-coupling regime; we need only to correct the baryon--$\gamma$ tight-coupling approximation in order to account for the new interaction. This can be easily achieved by following the step-by-step calculation of Ref.~\cite{Blas:2011rf}, including the additional terms $- \dot \mu (\theta_\gamma - \theta_{\rm{DM}})$ and $-S^{-1} \dot \mu (\theta_{\rm{DM}} - \theta_\gamma)$ in the photon and DM Euler equations respectively.

We implemented these modifications at order one in $\tau_c$ (and even beyond that order, since we used the approximation scheme called {\tt class\_compromise} in Ref.~\cite{Blas:2011rf}). We checked the consistency of our approach by varying the time at which the tight-coupling approximation is switched off in the presence of a non-zero interaction rate, $\dot \mu$. As expected, the results are independent of the switching time, unless it gets too close to recombination (in which case, one needs to introduce a DM--$\gamma$ tight-coupling regime).

In order to follow a reduced number of multipoles in the hierarchy of Boltzmann equations for photons, we expressed the final temperature and polarisation spectra using a line-of-sight integral \cite{Seljak:1996is}, i.e. we decompose the temperature/polarisation photon transfer functions $\Delta_l^{T,P}(k)$ as
\begin{equation}
\Delta_\ell^{T,P}(k) = \int_{\tau_i}^{\tau_0} d \tau~S^{T,P}\!(k,\tau)~j_\ell[k(\tau_0-\tau)]~,
\end{equation}
where $\tau$ is conformal time, $\tau_i$ is an arbitrary time much earlier than recombination, $\tau_0$ is the time today, $S^{T,P}(k,\tau)$ is the temperature/polarisation source function and the $j_\ell$'s are spherical Bessel functions. The source functions can be obtained by integrating the Boltzmann equation by parts along a given geodesic. For the model at hand, the source functions for temperature and polarisation read
\begin{eqnarray}
S^T \hspace{-1.1ex}&=&\hspace{-0.6ex}e^{-\kappa-\mu} \dot{\phi} + \frac{g}{4} \left(\delta_\gamma+ \frac{\Pi}{4}\right) + \frac{e^{-\kappa-\mu}}{k^2} \times
\nonumber \\
&&\hspace{-0.6ex}\left\{\left[\ddot \kappa + \dot \kappa(\dot \kappa + \dot \mu)\right] \theta_{\rm b} + \dot \kappa \dot \theta_{\rm b} + \left[\ddot \mu + \dot \mu(\dot \kappa + \dot \mu)\right] \theta_{\rm{DM}} + \dot \mu \dot \theta_{\rm{DM}} \right\} \nonumber \\
&&\hspace{-0.6ex}+ \frac{d}{d\tau} \left[e^{-\kappa-\mu} \psi + \frac{3}{16 k^2} (\dot g \Pi + g \dot \Pi) \right]~, \label{eq:S_T}\\
S^P\hspace{-1.1ex}&=&\hspace{-0.6ex}\frac{3}{16} \frac{g \Pi}{[k(\tau_0-\tau)]^2}~,
\end{eqnarray}
where $\Pi$ is a linear combination of temperature and polarisation multipoles, corresponding to $[F_{\gamma 2}+ G_{\gamma 0} + G_{\gamma 2}]$ in the notation of Ref.~\cite{Ma:1995ey}.

In the above formulae, for our numerical implementation in {\sc class}, derivatives of perturbations denoted with a dot are evaluated analytically using the evolution equations, while the derivative denoted by $\mathrm{d}/\mathrm{d}\tau$ is computed with a finite difference method, after storing the function between the square brackets.

\begin{figure*}[ht!]
\centering
\hspace{-3ex}
\includegraphics[width=8.8cm, trim = 2cm 1cm 1.3cm 1cm]{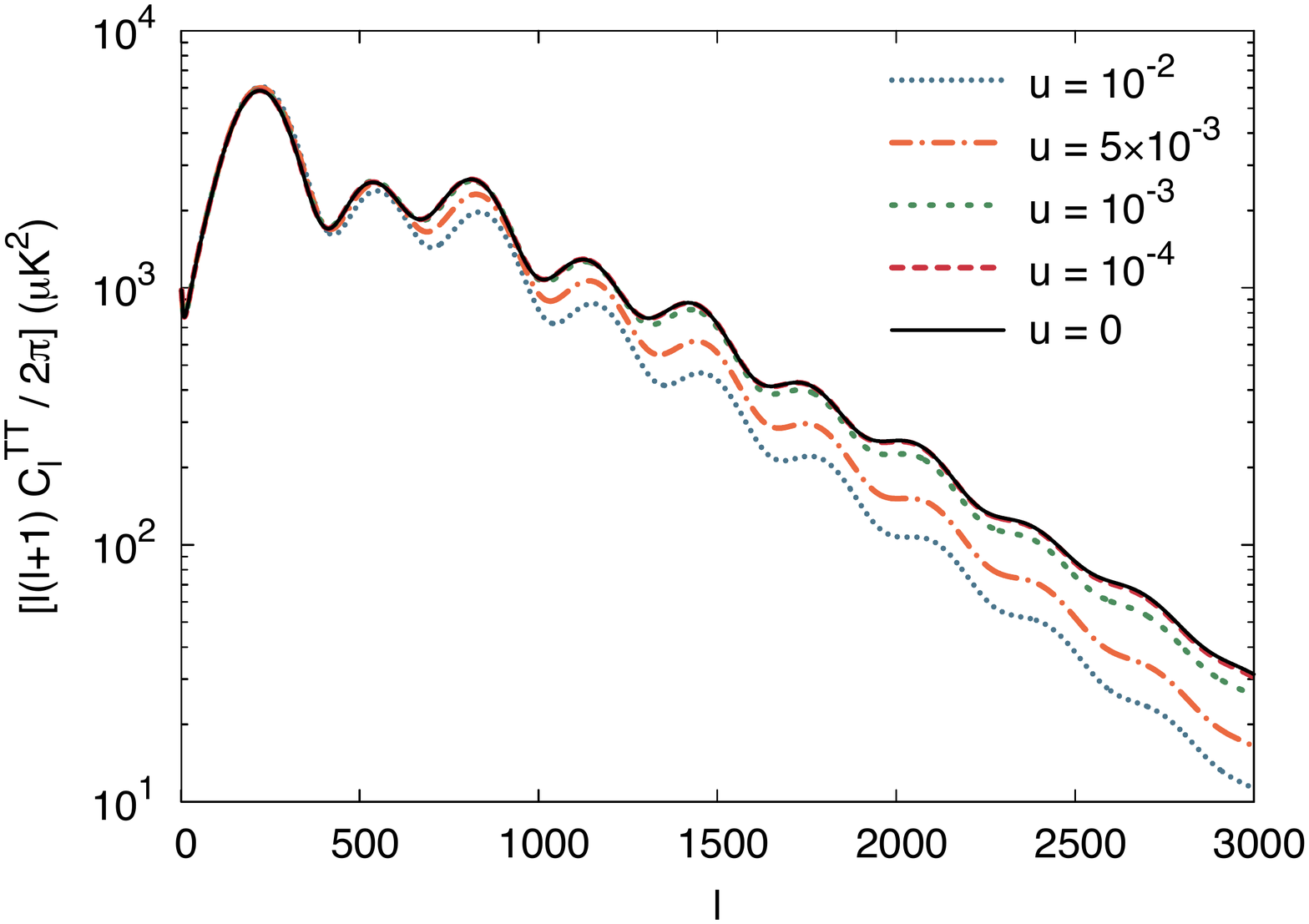}
\hspace{2ex}
\includegraphics[width=8.8cm, trim = 2.3cm 1cm 1cm 1cm]{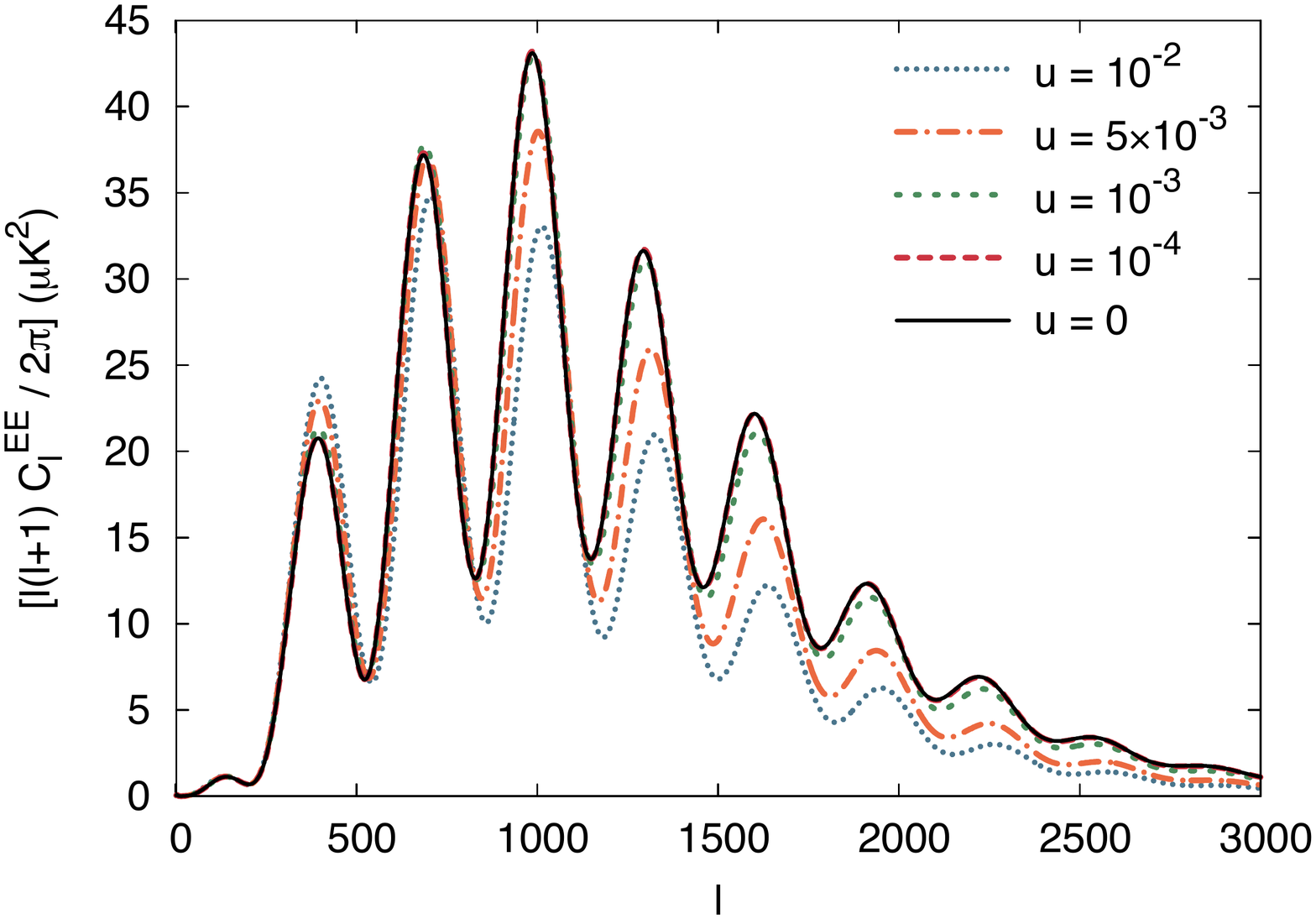}
\vspace{-3ex}
\caption{The effect of DM--$\gamma$ interactions on the $TT$ (left) and $EE$ (right) components of the CMB angular power spectrum, where the strength of the interaction is characterised by $u \equiv 
         \left[{\sigma_{\rm{DM}-\gamma}}/{\sigma_{\mathrm{Th}}} \right]
         \left[{m_{\rm{DM}}}/{100~\rm{GeV}} \right]^{- 1}$ ($u=0$ corresponds to zero DM--$\gamma$ coupling) and $\sigma_{\rm{DM}-\gamma}$ is constant. For all the curves, we consider a flat $\Lambda$CDM model with $H_0 = 70~\mathrm{km}~{\mathrm{s}}^{-1}~{\mathrm{Mpc}}^{-1}$ ($h = 0.7$), $\Omega_{\Lambda} = 0.7$, $\Omega_{\mathrm{m}} = 0.3$ and $\Omega_{\mathrm{b}} = 0.05$, where $u$ is the only additional parameter. The new coupling has two main effects: i) a suppression of the small-scale peaks due to a combination of collisional damping and a delayed photon decoupling, and ii) a shift in the peaks to larger $\ell$ due to a decrease in the sound speed of the thermal plasma. (Note that $u = 10^{-4}$ is difficult to distinguish from $u = 0$ at this scale).}
\label{fig:c_l}
\end{figure*}

%%%%%%%%%%%%%%%%%%%%%%%%%%%%%%%%%%%%%%%%%%%%%%%%%%%%%
\subsection{Effect of DM--$\gamma$ interactions on the CMB spectrum}
\label{subsec:cmb}
%%%%%%%%%%%%%%%%%%%%%%%%%%%%%%%%%%%%%%%%%%%%%%%%%%%%%

The impact of DM--$\gamma$ interactions on the $TT$ and $EE$ components of the CMB angular power spectrum generated by {\sc class} is illustrated in Fig.~\ref{fig:c_l} for specific values of the parameter $u \equiv \left[{\sigma_{\rm{DM}-\gamma}}/{\sigma_{\mathrm{Th}}} \right] \left[{m_{\rm{DM}}}/{100~\rm{GeV}} \right]^{- 1}$. Here we take the ${\rm{DM}-\gamma}$ cross section to be constant, however, we note that similar effects are observed for temperature-dependent cross sections.

For illustrative purposes, we consider a flat $\Lambda$CDM cosmology, where the energy content of the Universe is divided between baryons ($\Omega_{\rm{b}} = 0.05$), dark matter ($\Omega_{\mathrm{DM}} = 0.25)$ and dark energy in the form of a cosmological constant ($\Omega_{\Lambda} = 0.7$). We select a present-day value for the Hubble parameter of $H_0 = 70~\mathrm{km}~{\mathrm{s}}^{-1}~{\mathrm{Mpc}}^{-1}$ ($h = 0.7$) and a standard value of 3.046 for the effective number of neutrino species~\cite{Mangano:2005cc}.

There are two important effects on the relative amplitude and position of the Doppler peaks with respect to standard $\Lambda$CDM, both of which can be used to constrain the DM--$\gamma$ elastic scattering cross section.

Firstly, the DM--$\gamma$ interactions induce collisional damping (see Ref.~\cite{Boehm:2000gq,Boehm:2004th}), thus reducing the magnitude of the small-scale peaks and effectively cutting off the angular power spectrum at lower values of $\ell$. For very large cross sections, this effect is enhanced by a delay in the epoch of photon last-scattering, increasing the width of the last-scattering surface. Secondly, the presence of significant DM--$\gamma$ interactions decreases the sound speed of the thermal plasma~\cite{Boehm:2001hm}. Acoustic oscillations have a lower frequency, leading to a shift in the position of the Doppler peaks to larger $\ell$.

We note that there is a slight enhancement of the first acoustic peak with respect to $\Lambda$CDM ($\sim 0.1$\% in $TT$ and $\sim 0.3$\% in $EE$ for $u = 10^{-4}$) due to a decrease in the diffusion length of the photons.

As expected, these effects are enhanced for a larger cross section or a smaller DM mass (i.e. a greater number density of DM particles for the same relic density), corresponding to a larger value of $u$ and a later epoch of DM--$\gamma$ decoupling. Therefore, by fitting the $TT$ and $EE$ components of the CMB spectrum with cosmological data, one can constrain the value of $u$ and thus determine the maximal scattering cross section that is allowed for a given DM mass.

%%%%%%%%%%%%%%%%%%%%%%%%%%%%%%%%%%%%%%%%%%%%%%%%%%%%%
\section{Results and Outlook}
%%%%%%%%%%%%%%%%%%%%%%%%%%%%%%%%%%%%%%%%%%%%%%%%%%%%%

In this section, we present our constraints on the DM--$\gamma$ elastic scattering cross section, which is considered to be either constant or proportional to the temperature squared. We discuss important features of the temperature and polarisation spectra in the presence of DM--$\gamma$ interactions and outline prospects for future CMB experiments.

%%%%%%%%%%%%%%%%%%%%%%%%%%%%%%%%%%%%%%%%%%%%%%%%%%%%%
\subsection{Constraints from the Planck One-Year Data Release}
\label{subsec:results}
%%%%%%%%%%%%%%%%%%%%%%%%%%%%%%%%%%%%%%%%%%%%%%%%%%%%%

To fit our model to the data, we varied the parameters of the minimal flat $\Lambda$CDM cosmology, namely: the baryon density ($\Omega_{\rm{b}}h^2$), the dark matter density ($\Omega_{\rm{DM}}h^2$), the scalar spectral index ($n_s$), the primordial spectrum amplitude ($A_s$), the reduced Hubble parameter ($h$) and the redshift of reionisation ($z_{\rm{reio}}$), supplemented by the additional parameter characterising the DM--$\gamma$ interaction strength, $u \equiv \left[{\sigma_{\rm{DM}-\gamma}}/{\sigma_{\mathrm{Th}}} \right]\left[{m_{\rm{DM}}}/{100~\rm{GeV}} \right]^{- 1}$.

\begin{table*}
  \begin{tabular*}{0.95\textwidth}{@{\extracolsep{\fill} }c|ccccccc}
     \hline
   &   $100~\Omega_{\rm{b}} h^2$ & $\Omega_{\rm{DM}} h^2$  & $100~h$ &
     $10^{+9}~A_{s }$ & $n_s$ & $z_{\rm{reio}}$ & $10^{+4}~u$  \\ \hline 
     \vspace{-2.5ex}
     &&&&&&&\\
~~Best-fit~~ & 2.199  & 0.1195 & 67.57 & 2.189 & 0.9627 & 11.02 & $\simeq$ 0  \\ [1ex]
     \vspace{-3ex}
     &&&&&&&\\
 ~~Mean$~\pm~\sigma$~~ & $2.210_{-0.033}^{+0.029}$ & $0.1201_{-0.0029}^{+0.0028}$ &
     $67.6_{-1.3}^{+1.2}$ & $2.201_{-0.060}^{+0.054}$ &
     $0.9625_{-0.0080}^{+0.0076}$ & $11.2_{-1.2}^{+1.2}$ & $< 1.173$ \\[1.5ex]\hline 
     \vspace{-2ex}
     &&&&&&&\\
 ~~`{\it Planck} + WP' ~~ & $2.205_{-0.028}^{+0.028}$ & $0.1199_{-0.0027}^{+0.0027}$ &
     $67.3_{-1.2}^{+1.2}$ & $2.196_{-0.060}^{+0.051}$ &
     $0.9603_{-0.0073}^{+0.0073}$ & $11.1_{-1.1}^{+1.1}$ & $-$ \\ [1.5ex]
     \hline
   \end{tabular*}
\caption{Best-fit values and minimum credible intervals at the 68\% confidence level of the cosmological parameters set by Planck, with $u \equiv \left[{\sigma_{\rm{DM}-\gamma}}/{\sigma_{\mathrm{Th}}} \right] \left[{m_{\rm{DM}}}/{100~\rm{GeV}} \right]^{- 1}$ as a free parameter and a constant $\sigma_{\rm{DM}-\gamma}$. For comparison, `{\it Planck} + WP' are the 68\% limits taken from Ref.~\cite{Ade:2013zuv}. $\Omega_{\rm{b}} h^2$ is the baryon energy density, $ \Omega_{\rm{DM}} h^2$ is the dark matter energy density, $h$ is the reduced Hubble parameter, $A_s$ is the primordial spectrum amplitude, $n_s$ is the spectral index and $z_{\rm{reio}}$ is the reionisation redshift.}
\label{tab:params}
\end{table*}

\begin{figure*}
\centering
\includegraphics[scale=0.407]{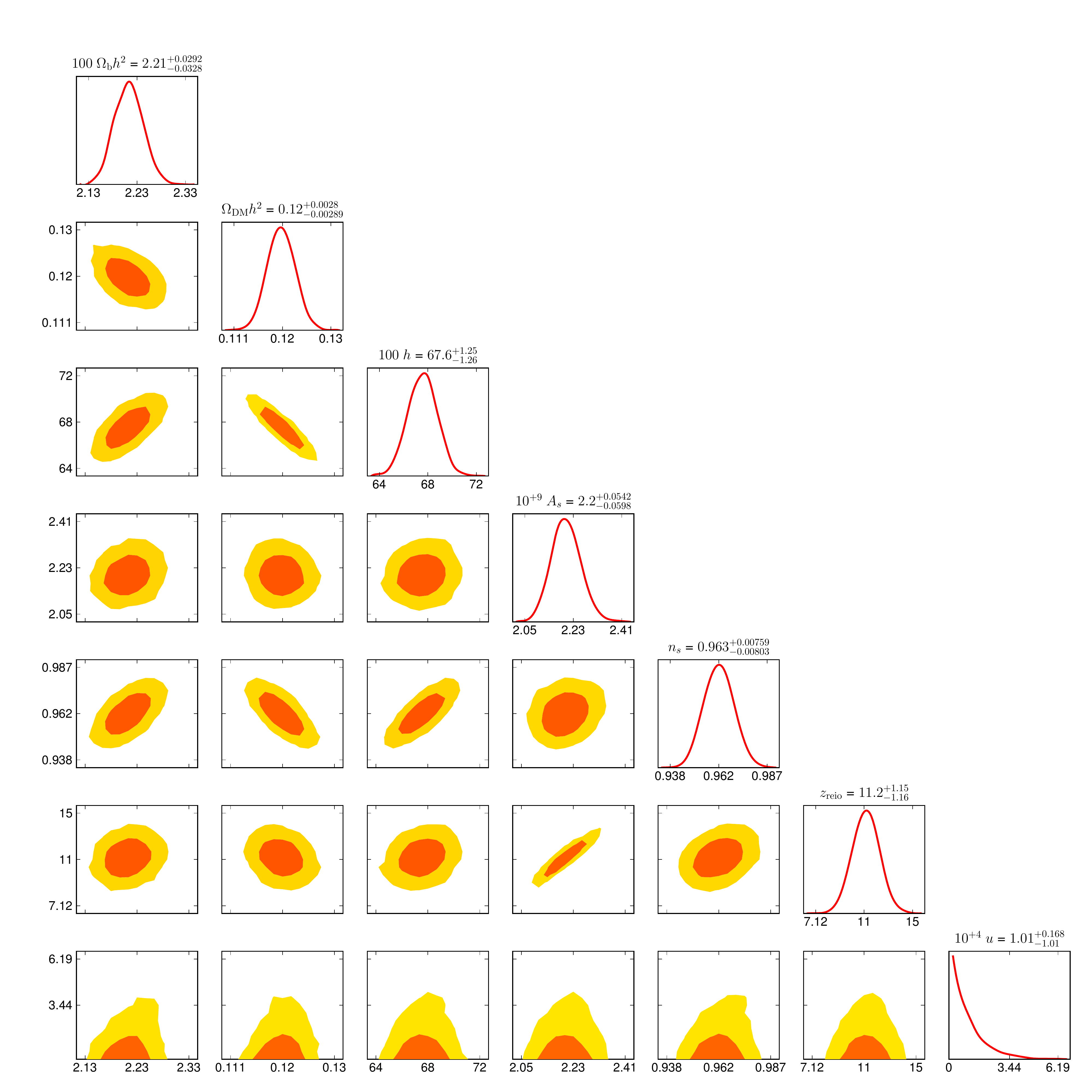}
\caption{Triangle plot showing the one and two-dimensional posterior distributions of the cosmological parameters set by Planck, with $u \equiv \left[{\sigma_{\rm{DM}-\gamma}}/{\sigma_{\mathrm{Th}}} \right] \left[{m_{\rm{DM}}}/{100~\rm{GeV}} \right]^{- 1}$ as a free parameter and a constant $\sigma_{\rm{DM}-\gamma}$. The contours correspond to the 68\% and 95\% confidence levels. $\Omega_{\rm{b}} h^2$ is the baryon energy density, $ \Omega_{\rm{DM}} h^2$ is the dark matter energy density, $h$ is the reduced Hubble parameter, $A_s$ is the primordial spectrum amplitude, $n_s$ is the spectral index and $z_{\rm{reio}}$ is the reionisation redshift.}
\label{fig:planck_fit}
\end{figure*}

We considered three active neutrino species; two massless and the other with a small mass of 0.06 eV, reflecting the lower bound imposed by neutrino oscillation experiments\footnote{This is an approximation that is used throughout literature, including in the Planck analysis~\cite{Ade:2013zuv}. Since the data is mainly sensitive to the sum of the neutrino masses~\cite{Lesgourgues:2006nd} and it is faster to run a Boltzmann code with only one massive neutrino, we also use this excellent approximation.}~\cite{Beringer:1900zz}. In addition, we chose the standard value of 3.046 for the effective number of neutrino species, $N_{\rm eff}$~\cite{Mangano:2005cc} (allowing $N_{\rm eff}$ to vary does not have a significant effect on our conclusions).

To efficiently sample the parameter space, we ran the Markov Chain Monte Carlo code {\sc Monte Python}\footnote{\tt montepython.net}~\cite{Audren:2012wb} combined with the one-year data release from Planck, provided by the Planck Legacy Archive\footnote{\tt  pla.esac.esa.int/pla/aio/planckProducts.html}~\cite{Ade:2013ktc}. In particular, we used the high-$\ell$ and low-$\ell$ temperature data of Planck combined with the low-$\ell$ WMAP polarisation data (this corresponds to `{\it Planck} + WP' in Ref.~\cite{Ade:2013zuv}). We marginalised over the nuisance parameters listed in Ref.~\cite{Ade:2013zuv}.

For a constant cross section, the bounds on the various cosmological parameters are displayed in Table~\ref{tab:params} and illustrated in Fig.~\ref{fig:planck_fit} (where we omit the nuisance parameters for clarity). The posterior probability distribution for the $u$ parameter peaks at $u\simeq 0$ showing that the data does not prefer a significant DM--$\gamma$ coupling. Importantly, we derive an upper limit on the elastic scattering cross section of
\begin{equation}
\sigma_{\rm{DM}-\gamma} \leq~8 \times 10^{-31} \left(m_{\rm{DM}}/\rm{GeV}\right) \ \rm{cm^2}~,
\label{eq:bound}
\end{equation}
corresponding to $u \leq 1.2 \times 10^{-4}$ (at 68\% CL). This result constitutes an improvement by a factor of $\sim$ 9 on the pre-WMAP analysis of Ref.~\cite{Boehm:2001hm}, which set a limit of $\sigma_{\rm{DM}-\gamma} \lesssim~7 \times 10^{-30} \left(m_{\rm{DM}}/\rm{GeV}\right) \ \rm{cm^2}$ by comparing the CMB anisotropy spectra with $\Lambda$CDM predictions.

We note that including data from the 2500-square degree SPT survey~\cite{Hou:2012xq} tightens the constraints on the cosmological parameters with respect to `{\it Planck} + WP' alone, giving best-fit values that are consistent at the 1$\sigma$ level. We obtain the slightly weaker result of $u \leq 1.3 \times 10^{-4}$, in addition to a larger value of $H_0=67.9_{-1.1}^{+1.0}$ and smaller value of $z_{\rm reio} = 10.7_{-1.2}^{+1.0}$ (all at 68\% CL)\footnote{Note that these results must be considered with care, given the small tension between the amplitudes of the CMB damping tail in the SPT and Planck data (as reported in Ref.~\cite{Ade:2013zuv}, although the Planck collaboration now has a better understanding of the source of this tension).}.

For a DM candidate that is lighter than a few GeV (see e.g.~\cite{Boehm:2002yz,Boehm:2003bt}), Eq.~\eqref{eq:bound} suggests that the particles must have a cross section in the range of weak interactions. This result is relevant for scenarios in which DM cannot annihilate directly into the visible sector (where indirect detection techniques are inappropriate). Meanwhile, for a heavy DM particle ($\sim$ TeV), we obtain a weaker bound on the scattering cross section so large DM--$\gamma$ interactions (with respect to weak interactions) cannot yet be ruled out by CMB data alone.

For scenarios where DM cannot couple directly to photons, this translates into an upper bound on the DM coupling to charged particles, including those of the SM. However, the requirement of a constant cross section implies that there is some cancellation that enables us to remove the dependence on the photon energy, as in the case of Thomson scattering. Scenarios in which the DM mass is degenerated with the mediator mass may therefore be more appropriate, providing that the mass degeneracy passes the cuts at the LHC (e.g. Ref.~\cite{Arbey:2013aba}) or the DM mass is large enough to satisfy the LHC constraints on new charged particles.

A constant cross section is also expected in the presence of a $Z'$--$\gamma$ or $\gamma'$--$\gamma$ mixing (for a review on the limits of such a mixing, see for example, Ref.~\cite{Jaeckel:2010ni}). In this case, the cross section is essentially the Thomson cross section (where we replace the fine structure constant, $\alpha$, by its equivalent for the DM-$\gamma'$ coupling, $\alpha_{\rm{DM}-\gamma'}$, and the electron mass by the DM mass) multiplied by the $Z'/\gamma'-\gamma$ coupling, $\chi$, to the power four (i.e. $\sigma_{\rm{DM}-\gamma} = \chi^4 \ \sigma_{\rm{DM}-\gamma'}$). For MeV DM, our result translates into the constraint $\chi \lesssim 10^{-2}$ in the limit of a massless $Z'$/$\gamma'$ and $\alpha_{\rm{DM}-\gamma'} \simeq \alpha$. This is to be compared with the bounds on millicharged particles, which are about two to three orders of magnitude stronger in the MeV range~\cite{Prinz:1998ua}.

If instead, the cross section is proportional to the temperature squared, we obtain the stringent upper bound of
\begin{equation}
\sigma_{\rm{DM}-\gamma} \leq~6 \times 10^{-40} \left(m_{\rm{DM}}/\rm{GeV}\right) \ \rm{cm^2}~,
\end{equation}
for the present-day value of the scattering cross section (at 68\% CL), corresponding to $u_0 \leq 9.0 \times 10^{-14}$, which is consistent with Eq.~\eqref{eq:bound}. Brought back to epochs much earlier than the CMB time, this result is clearly not as powerful as the constant cross section case, but does apply to all scenarios where the dependence on the energy of the photon cannot be alleviated. In the case of dipolar DM models~\cite{Sigurdson:2004zp,Gardner:2008yn,Chang:2010en}, this enables one to constrain the DM dipole moment.

We note that our analysis assumes that the interacting DM accounts for the entire DM component of the Universe; if more than one species were responsible for the observed relic density \cite{Boehm:2003ha}, larger cross sections would be allowed.

%%%%%%%%%%%%%%%%%%%%%%%%%%%%%%%%%%%%%%%%%%%%%%%%%%%%%
\subsection{Prospects for future experiments}
\label{subsec:predictions}
%%%%%%%%%%%%%%%%%%%%%%%%%%%%%%%%%%%%%%%%%%%%%%%%%%%%%

As shown in Table~\ref{tab:params}, our best fit to the Planck data for $u \lesssim 10^{-4}$ leads to values of the cosmological parameters that are consistent with those obtained by Planck at the 1$\sigma$ level. However, there are a number of differences with respect to $\Lambda$CDM at high $\ell$ due to the impact of DM--$\gamma$ interactions, which eventually dampen structure on very small scales.

\begin{figure*}
\centering
\hspace{-3ex}
\includegraphics[width=8.8cm, trim = 2cm 1cm 1.3cm 1cm]{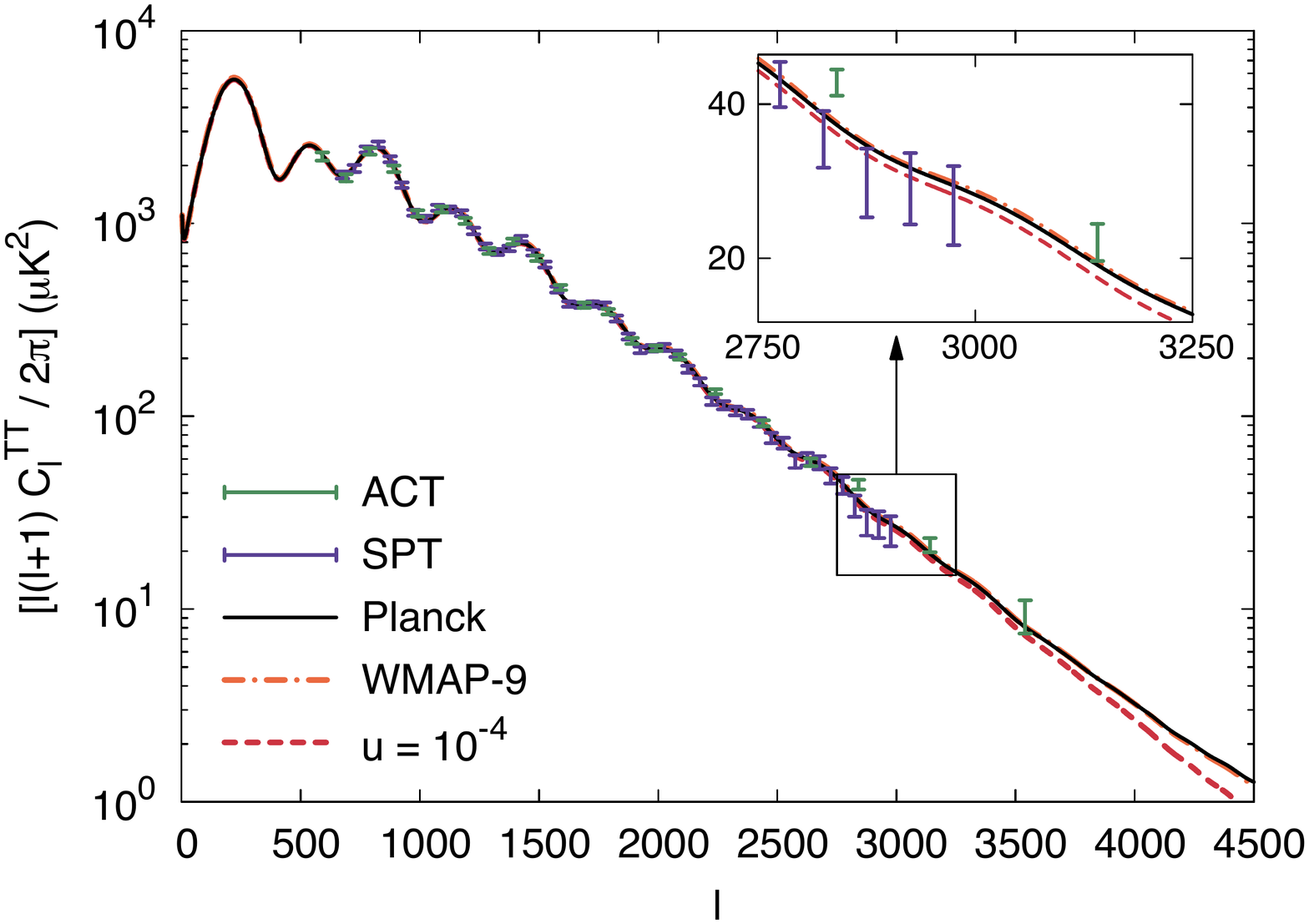}
\hspace{1ex}
\includegraphics[width=8.8cm, trim = 2.3cm 1cm 1cm 1cm]{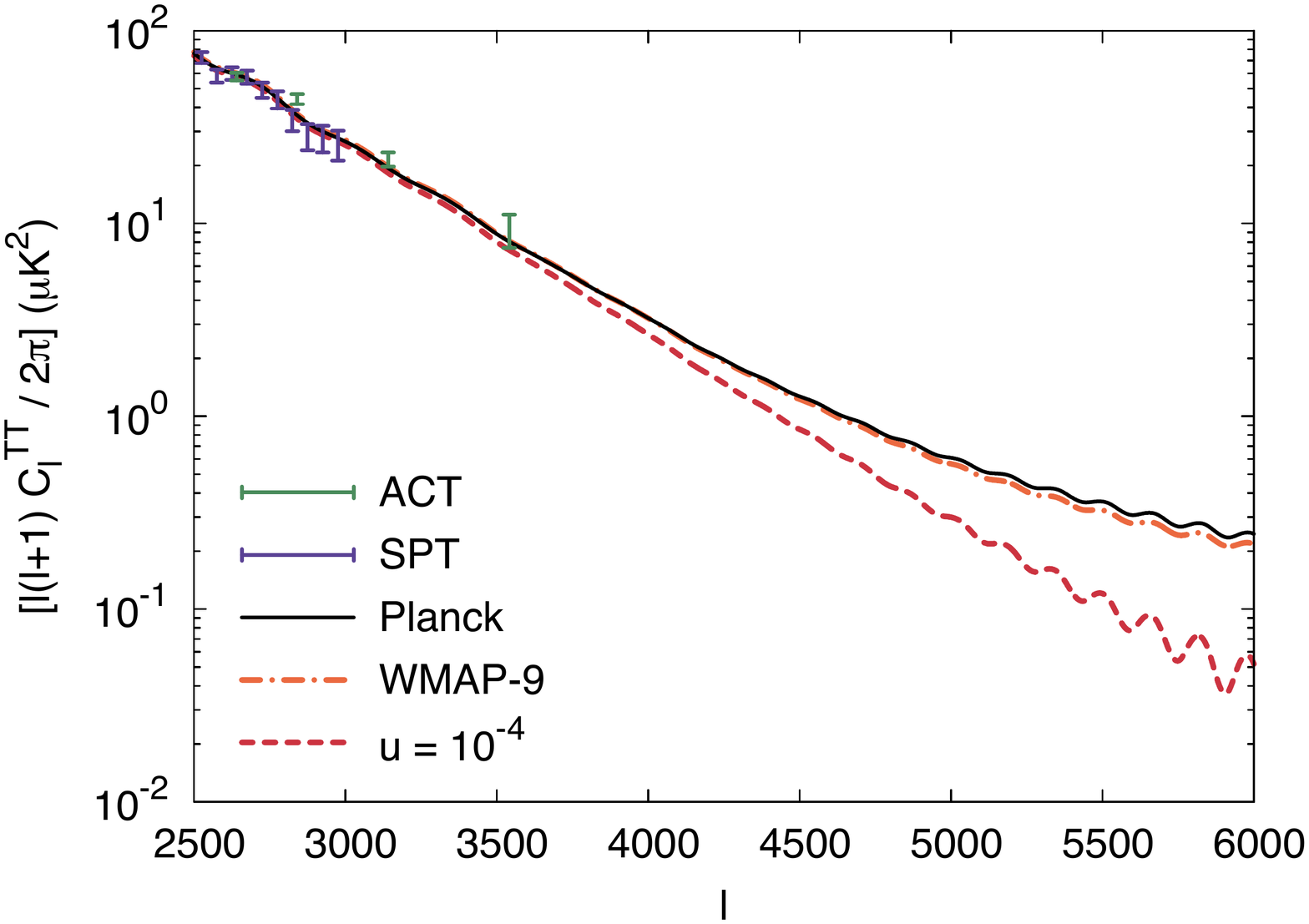}
\vspace{-3ex}
\caption{A comparison between the $TT$ angular power spectra for the maximally allowed (constant) DM--$\gamma$ cross section ($u \simeq 10^{-4}$), and the 9-year WMAP~\cite{Hinshaw:2012aka} and one-year Planck~\cite{Ade:2013zuv} best-fit data. Also plotted are the full 3-year data from the SPT and ACT experiments~\cite{Calabrese:2013jyk}. On the left, we see a suppression of power with respect to WMAP-9 and Planck for $\ell \gtrsim 3000$ and on the right, we give our prediction for the $TT$ component of the angular power spectrum at high $\ell$.}
\vspace{-4ex}
\label{fig:c_l_data}
\end{figure*}

\begin{figure}
\centering
\includegraphics[width=8.8cm, trim = 3cm 2.5cm 0.3cm 1cm]{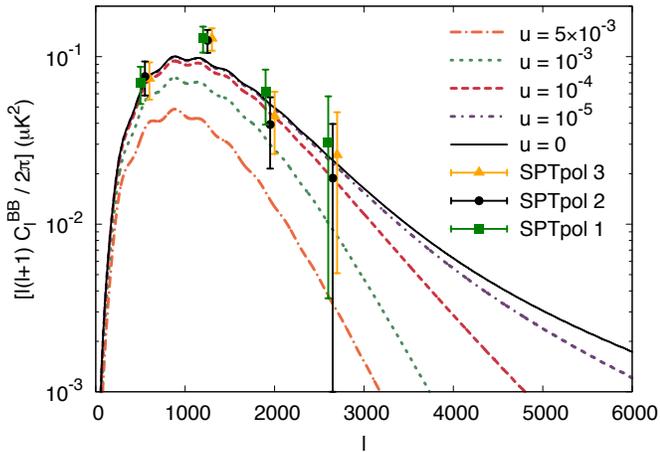}
\caption{The effect of DM--$\gamma$ interactions on the $B$-modes of the angular power spectrum, where the strength of the interaction is characterised by $u \equiv 
         \left[{\sigma_{\rm{DM}-\gamma}}/{\sigma_{\mathrm{Th}}} \right]
         \left[{m_{\rm{DM}}}/{100~\rm{GeV}} \right]^{- 1}$ (with a constant $\sigma_{\rm{DM}-\gamma}$) and we use the `{\it Planck} + WP' best-fit parameters from Ref.~\cite{Ade:2013zuv}. The data points are the recent $B$-mode polarisation measurements from the SPT experiment, where SPTpol 1, SPTpol 2 and SPTpol 3 refer to $({\hat{\rm{E}}}^{150}{\hat{\phi}}^{\rm{CIB}}) \times {\hat{\rm{B}}}^{150}$, $({\hat{\rm{E}}}^{95}{\hat{\phi}}^{\rm{CIB}}) \times {\hat{\rm{B}}}^{150}$ and $({\hat{\rm{E}}}^{150}{\hat{\phi}}^{\rm{CIB}}) \times {\hat{\rm{B}}}_{\chi}^{150}$ respectively in Ref.~\cite{Hanson:2013hsb}. For the maximally allowed (constant) DM--$\gamma$ cross section ($u \simeq 10^{-4}$), we see a deviation from the Planck best-fit $\Lambda$CDM model for $\ell \gtrsim 500$ and a significant suppression of power for larger $\ell$.}
\vspace{-0.5ex}
\label{fig:bb}
\end{figure}

\begin{figure}
\centering
\includegraphics[width=8.8cm, trim = 3cm 2.5cm 0.3cm 1cm]{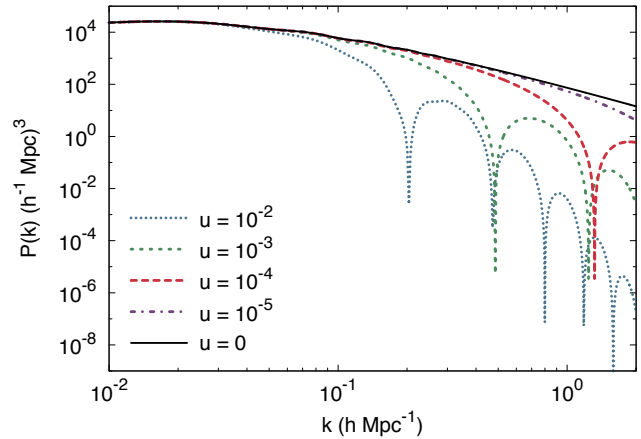}
\caption{The influence of DM--$\gamma$ interactions on the matter power spectrum, where the strength of the interaction is characterised by $u \equiv 
         \left[{\sigma_{\rm{DM}-\gamma}}/{\sigma_{\mathrm{Th}}} \right]
         \left[{m_{\rm{DM}}}/{100~\rm{GeV}} \right]^{- 1}$ (with a constant $\sigma_{\rm{DM}-\gamma}$) and we use the `{\it Planck} + WP' best-fit parameters from Ref.~\cite{Ade:2013zuv}. The new coupling produces (power-law) damped oscillations at large scales, reducing the number of small-scale structures, thus allowing the cross section to be constrained. For allowed (constant) DM--$\gamma$ cross sections ($u \lesssim 10^{-4}$), significant damping effects are restricted to the non-linear regime ($k \gtrsim 0.2~h~{\rm{Mpc}}^{-1}$).}
\vspace{2.5ex}
\label{fig:p(k)}
\end{figure}

The effect is particularly noticeable if one considers the $TT$ angular power spectrum for $\ell \gtrsim 3000$, which has not yet been probed by Planck (see Fig.~\ref{fig:c_l_data}). Indeed, for $\ell \simeq 6000$, small-scale fluctuations are suppressed by a factor of $\sim$ 4 with respect to $\Lambda$CDM for our maximally allowed cross section. This result could be promising for CMB experiments such as  SPT~\cite{Hou:2012xq} and ACT~\cite{Sievers:2013ica}; however, such a large value of $\ell$ corresponds to the region where the foregrounds (emission from extra-galactic sources and the thermal Sunyaev-Zeldovich effect) are dominant~\cite{Crawford:2013uka}. Therefore, the detectability of DM--$\gamma$ interactions in the temperature anisotropy spectrum will depend on the accuracy of foreground modelling and removal.

The damping with respect to $\Lambda$CDM is also evident in the $B$-mode spectrum (a consequence of $E$-mode lensing by large-scale structure), as shown in Fig.~\ref{fig:bb}. The reduction in power is due to the combined damping of the $E$-modes (see Fig.~\ref{fig:c_l}) and the matter power spectrum (see Fig.~\ref{fig:p(k)}). While the overall effect is small for $u \lesssim 10^{-4}$, if we consider $\ell \gtrsim 500$, one can use the $B$-modes alone combined with the first-season SPTpol data~\cite{Hanson:2013hsb} to effectively rule out $u \gtrsim 5 \times 10^{-3}$. In fact, future polarisation data from e.g. SPT~\cite{Hou:2012xq}, POLARBEAR~\cite{Kermish:2012eh} and SPIDER~\cite{Crill:2008rd} could be sensitive enough to distinguish $u \simeq 10^{-5}$ from $\Lambda$CDM.

Finally, the matter power spectrum may provide us with an even stronger limit on the DM--$\gamma$ interaction cross section (see Fig.~\ref{fig:p(k)}). The pattern of oscillations together with the suppression of power at small scales, as noticed already in Ref.~\cite{Boehm:2001hm}, could indeed constitute an interesting signature. The observability of such an effect depends on the non--linear evolution of the matter power spectrum (for which $k \gtrsim 0.2~h~{\rm{Mpc}}^{-1}$). Typically, one would expect it to be somewhat intermediate between cold and warm dark matter (WDM) scenarios at large redshifts, and closer to WDM at small redshifts so the Lyman-$\alpha$ constraint on WDM models could apply. Using the latest bound on the mass of WDM candidates~\cite{Viel:2013fqw} together with the proposed transfer function in Ref.~\cite{Boehm:2001hm}, we expect structure formation to set a more stringent limit than CMB analysis (potentially by several orders of magnitude) but this would require a thorough investigation.

Given that simulating non-linear structure formation in WDM models is renowned to be very challenging (due to numerical artefacts that are difficult to remove), we expect that constraining DM--$\gamma$ interactions through their matter power spectrum and distinguishing their effects from those of collisionless WDM will not be straightforward. Nevertheless, simulating such oscillating power spectra would enable us to study the impact of DM--$\gamma$ interactions in the non-linear regime, thereby determining the number of substructures for these models. This would be particularly useful in light of forthcoming data from large-structure surveys such as SDSS-III~\cite{Eisenstein:2011sa} and Euclid~\cite{Laureijs:2011gra}.

%%%%%%%%%%%%%%%%%%%%%%%%%%%%%%%%%%%%%%%%%%%%%%%%%%%%%
\section{Conclusion}
\label{sec:conc}
%%%%%%%%%%%%%%%%%%%%%%%%%%%%%%%%%%%%%%%%%%%%%%%%%%%%%

We have studied the effects of introducing an interaction between DM and photons on the evolution of primordial matter fluctuations and in particular, the CMB temperature and polarisation power spectra. By comparing the $TT$ and $EE$ components in the presence of a DM--$\gamma$ coupling with the one-year data release from Planck, we have set a stringent constraint on the elastic scattering cross section of $\sigma_{\rm{DM}-\gamma} \leq~8 \times 10^{-31} \left(m_{\rm{DM}}/\rm{GeV}\right) \ \rm{cm^2}$ (68\% CL), assuming it is constant at late times. This bound is an order of magnitude stronger than the previous work of Ref.~\cite{Boehm:2001hm}, where a limit was placed by comparing the temperature anisotropy spectrum with $\Lambda$CDM predictions (before any experimental results were published).

For a heavy DM particle ($\sim$ TeV), the maximal cross section is too large to exclude the possibility that DM has significant interactions with photons, while for light DM particles ($\sim$ MeV), the cross section is of the order typically expected for weak interactions. If instead, the cross section is proportional to the temperature squared, we obtain a significantly tighter present-day bound of $\sigma_{\rm{DM}-\gamma} \leq~6 \times 10^{-40} \left(m_{\rm{DM}}/\rm{GeV}\right) \ \rm{cm^2}$ (68\% CL), giving a weaker constraint in the past.

We note that an even stronger result could be achieved using forthcoming data on the $B$-modes and measurements of the $TT$ spectrum at very high $\ell$, provided an excellent knowledge of the foregrounds. We expect these limits to be weaker than those from the matter power spectrum, when combined with data from large-scale structure surveys and Lyman-$\alpha$ constraints (at present, a limit can only be set by analogy with collisionless WDM). However, CMB constraints will be important to compare to since they do not depend on the non-linear evolution of the matter fluctuations.

Importantly, we have shown that one can effectively use cosmological data to restrict the allowed region of parameter space for DM interactions, independently of any theoretical prejudice. Indeed, any CMB experiment with the ability to measure the power spectra at high $\ell$ could contribute to our fundamental understanding of DM interactions.

%%%%%%%%%%%%%%%%%%%%%%%%%%%%%%%%%%%%%%%%%%%%%%%%%%%%%
\acknowledgements
%%%%%%%%%%%%%%%%%%%%%%%%%%%%%%%%%%%%%%%%%%%%%%%%%%%%%

The authors would like to thank Carlton Baugh, Carlos Frenk, Martin Haehnelt, Jonathan Davis, Erminia Calabrese, Joerg Jaeckel and Silvia Pascoli for useful discussions, Benjamin Audren for his assistance with {\sc Monte Python}, and the Planck Editorial Board and Science Team for their reading of this work. We acknowledge the use of the publicly available numerical codes {\sc class} and {\sc Monte Python}, and the CMB data from the WMAP, Planck, SPT and ACT experiments. RJW and CB are supported by the STFC. This work was supported in part by the National Science Foundation under Grant No. PHYS-1066293, and the hospitality of the Aspen Center for Physics and the European Union FP7 ITN INVISIBLES.

%%%%%%%%%%%%%%%%%%%%%%%%%%%%%%%%%%%%%%%%%%%%%%%%%%%%%
%%%%%%%%%%%%%%%%%%%%%%%%%%%%%%%%%%%%%%%%%%%%%%%%%%%%%

\bibliography{DMGAMMA.bib}

\end{document}